\documentclass[11pt, amsmath, amssymb]{revtex4-1}%{elsarticle}
\usepackage{bm,amsfonts, mathtools}
\usepackage{graphicx,color, wasysym}
\usepackage{textcomp}
\usepackage{amsmath,amssymb,latexsym,epsfig}
\usepackage[british]{babel}
\usepackage{hyphenat}
\usepackage{appendix}

\def\ulamek#1#2{\mbox{\normalfont$\frac{#1}{#2}$}}

\DeclareMathOperator{\okr}{{\stackrel{{\scriptscriptstyle{\mathsf{def}}}}{=}}}

\DeclareMathOperator{\D}{d\!}
\DeclareMathOperator{\E}{e} 
\DeclareMathOperator{\I}{i}
   \DeclareMathOperator{\RE}{\mathfrak{Re}}

\newtheorem{proposition}{Proposition}

\newtheorem{example}{Example}

\begin{document} 

%\allowdisplaybreaks

%\makeatletter

\title{The Volterra type equations related to the non-Debye relaxation.}

\author{K.~G\'{o}rska}
\email{katarzyna.gorska@ifj.edu.pl}
\author{A. Horzela}
\email{andrzej.horzela@ifj.edu.pl}
\affiliation{H. Niewodnicza\'{n}ski Institute of Nuclear Physics, Polish Academy of Sciences, \\ 
ul.Eljasza-Radzikowskiego 152, PL 31342 Krak\'{o}w, Poland}

\begin{abstract}
We investigate a possibility to describe the non-Debye relaxation processes using the Volterra-type equations with kernels given by the Prabhakar functions with the upper parameter $\nu$ being negative. Proposed integro-differential equations mimic the fading memory effects and are explicitly solved using the umbral calculus and the Laplace transform methods. Both approaches lead to the same results valid for admissible domain of the parameters $\alpha$, $\mu$ and $\nu$ characterizing the Prabhakar function. For the special case $\alpha\in (0,1]$, $\mu=0$ and $\nu=-1$  we recover the Cole-Cole model, in general having a residual polarization.  We also show that our scheme gives results equivalent to those obtained using the stochastic approach to relaxation phenomena merged with integral equations involving kernels given by the Prabhakar functions with the positive upper parameter.  
\end{abstract}


\maketitle

\section{Introduction}

Dielectric relaxation phenomena which take place in dielectric materials reflect delay of the molecular polarization with respect to the time varying electric field acting on a dielectric medium. They are customarily described in terms of the frequency dependent complex dielectric permittivity ${\hat\varepsilon}(\I\!\omega)$ which informs us  about the reaction of a system on harmonic field and  is specified by a complex function whose real and imaginary parts are associated with absorptive and dispersive contributions. Our knowledge of these functions comes from the broadband dielectric spectroscopy and as such it has phenomenological origin which awaits theoretical clarification. In the case of an ideal, noninteracting population of dipoles its response to an alternating external electric field, in the frequency domain described by the so-called spectral function ${\hat\phi(\I\!\omega)} =  [{\hat\varepsilon}(\I\!\omega) - \varepsilon_{\infty}]/(\varepsilon_{0} - \varepsilon_{\infty})$, is indicated by the Debye relaxation, namely  ${\hat\phi(\I\!\omega)} = [1 + (\I\!\omega \tau)]^{-1}$ where $\tau$ denotes some characteristic time. In the time domain the Debye relaxation pattern can be described by the relaxation function $f(t) = \exp(-t/\tau)$ whose derivative taken with minus sign gives the response function $\phi (t) = \mathcal{L}^{-1}[{\hat{\phi}(\I\!\omega)};t]$ where $\mathcal{L}^{-1}[\cdot\,;\; t]$ denotes the inverse Laplace transform. For more complex systems the Debye model fails and the differential equation of the Debye relaxation function, i.e. $\D f(t)/\!\D t = -\tau^{-1} f(t)$, is not satisfied any longer. 
{Intuitive explanation says that the system does not follow either changes or switching on/off the external field immediately and its observed reaction is delayed and smeared in time.} {In the language of evolution equations such time-smearing  may  be written down in various forms.  Within the approach proposed in  \cite{AAKhamzin14, AStanislavsky17, AStanislavsky18, AStanislavsky19} the starting point is the kinetic (master) equation $\D f(t)/\D t = - r(t) f(t)$ with a non-negative function  $r(t)$ called the transition rate of the system. The master equation may be rewritten as $f(t) = f(0) - \int_{0}^{t} r(\xi) f(\xi) \D\xi$ where $f(0) =1$. Smearing the function $r(\xi)$ as $r(t-\xi) = B(\tau;\textbf{p}) \kappa(t-\xi)$ where $\textbf{p}$ denote a set of external parameters characterizing the system, we arrive at  the inhomogeneous integral equation  
\begin{equation}\label{9/02-1}
f(t) = 1 - B(\tau;\textbf{p}) \int_{0}^{t} \kappa(t - \xi) f(\xi) \D\xi,
\end{equation}
recently considered in detail within the framework  of the stochastic approach to relaxation processes \cite{AStanislavsky17, AStanislavsky18, AStanislavsky19}. In Eq. \eqref{9/02-1} the kernel $\kappa(t)$ plays the role of propagator and the quantity $B(\tau;\textbf{p})$ (reducing to $1/\tau$ for the Debye relaxation) contains parameters which characterize the process under consideration. Another possibility is to smear the time derivative $\frac{\D f(t)}{\D t}$. It leads to the homogenous integro-differential equation 
}
\begin{equation}\label{12/03-2}
\int_{0}^{t} k(t-\xi) \frac{\D}{\D\xi} f(\xi) \D\xi = - B(\tau; \textbf{p}) f(t),
\end{equation}
 with the kernel $k(t)$ which mimics memory effects influencing the system evolution and thus we shall call it the memory kernel. Suitable choice of $k(t)$ enables using fractional analysis tools being nowadays widespread mathematical methods applied to investigate various kinds of relaxation and anomalous diffusion phenomena. We note here that the functional form of $k(t)$ depends on the physical model under study and is by no means {\sl {a priori}} restricted to the forms present in the standard definitions of fractional derivatives, like the Riemann-Liouville or Caputo ones. It has been advocated in %important papers 
Refs. \cite{ECdOliveira19,RGarra14,RGarrappa16,AAKhamzin14,Nigmatullin97,AStanislavsky18} that the promising and physically justified choice of the memory kernel is to take it as the Prabhakar function \cite{TRPrabhakar71}, being a representative of the Mittag-Leffler family of functions \cite{Mainardi-b-ML} playing important role in modeling the non-Debye relaxations. In this paper, following arguments presented in \cite{ECdOliveira19,Mainardi-b-ML}, we shall focus our  interest  on the memory kernel given by the Prabhakar function with negative upper parameters, namely
\begin{equation}\label{5/04-1}
{k(\alpha, \nu, \mu; a; t) = e^{-\nu}_{\alpha, \,1 - \mu}(-a, t) = t^{-\mu} E^{-\nu}_{\alpha, \,1 - \mu}(-a t^{\alpha}), \qquad a\in \mathbb{R}, \quad t > 0,}
\end{equation} 
with $\alpha, \nu, \mu \in(0, 1]$. 

The Eq. \eqref{12/03-2}, with Eq. \eqref{5/04-1} and $B(\tau;\alpha, \nu, \mu)$ put in, is an integro-differential equation of the Volterra type. We will solve it using two independent techniques namely the umbral calculus and the Laplace transform methods. Tricky methods of the umbral calculus simplify calculations leading to the main results of the paper. However, because umbral calculus methods are repeatedly considered non reliable all results are confirmed using the standard Laplace transform method. {Having these results we identify examples of relaxation processes for which Eq. \eqref{12/03-2} provide us with physically adequate description. Finally, we show that the inhomogeneous integral equation Eq. \eqref{9/02-1} is equivalent to the homogenous integro-differential equation Eq. \eqref{12/03-2}.}

The paper is organized as follows: in the Sec. \ref{sec1} we recall solvability conditions for Eq. \eqref{12/03-2} (obtained in \cite{Kochubei2011}) as well as discuss the series form of its solutions in the Laplace space. The Secs. \ref{sec2} and \ref{sec3} are a mathematical {\em interlude}. We have decided to include them in order to make the paper self-consistent and to convince the reader to usability of the umbral methods. In Sec. \ref{sec2} we introduce the umbral representation of the three parameter Mittag-Leffler function $E^{\nu}_{\alpha, \mu}(x)$ and show that the umbral techniques permit one to reconstruct numerous formulas obeyed by the Mittag-Leffler functions. Analogous scheme is repeated in Sec. \ref{sec3} for the Mittag-Leffler polynomials. The Secs. \ref{sec4} and \ref{sec6} are devoted to study explicitly solvable examples to which the equation Eq. \eqref{12/03-2} boils down: in Sec. \ref{sec4} we solve it for the kernel $k(\alpha, 1; a, t) = e^{-1}_{\alpha, 1 - \alpha}(-a, t)$, $\alpha\in(0, 1)$; whereas in Sec. \ref{sec6} we consider the kernel $k(\alpha, \nu; a, t) = e^{-\nu}_{\alpha, 1 - \nu\alpha}(-a, t)$, $\alpha, \nu\in (0, 1]$. We show how to sum up the series expressing the inverse Laplace transform (giving solutions to Eqs. \eqref{21/03-1} and \eqref{21/03-2}) by applying the umbral calculus and the Laplace transform methods. Both results are equivalent and valid in the full range of parameters so there is no need to assume any longer restrictions initially requested by Eqs. \eqref{21/03-1} and \eqref{21/03-2}. In Sec. \ref{sec7} we consider the case of $k(\alpha, \nu, \mu; a, t) = e^{-\nu}_{\alpha, 1 - \mu}(-a, t)$, $\alpha, \nu, \mu\in (0, 1]$ which, being depended  on three parameters, generalizes results of Sec. \ref{sec6} and {it} is closely related to the example analysed in \cite{AStanislavsky18}. The solution which we have found involves unknown inverse Laplace transform and gives, as a special case, the solution presented in Sec. \ref{sec6}. In Sec. \ref{sec8} we conclude the paper - in particular we show how our results are related to the Cole-Cole relaxations and how they link to parallel investigations based on integral equations and stochastic approach.

\section{Solutions to the Volterra equation: the Laplace transform methods}\label{sec1}

Eq. \eqref{12/03-2} defines the Cauchy problem which is well understood under quite general assumptions put on the kernel $k(\alpha, \nu, {\mu}; a; t)$ \cite{Kochubei2011}. Properties of  $k(\alpha, \nu, {\mu}; a; t)$ are the best recognized if one describes them in terms of its Laplace transform ${\hat{k}}(\alpha, \nu, {\mu}; a; s)$.  It has been shown in \cite[Theorem 2]{Kochubei2011} that the solvability of Eq. \eqref{12/03-2} is determined by the conditions
\begin{equation}\label{17_08_1}
\begin{array}{rcl}
{\hat{k}}(\alpha, \nu, {\mu}; a; s)\to\infty,&\quad\quad s{\hat{k}}(\alpha, \nu, {\mu}; a; s)\to 0,\quad&\text{ for}\quad\quad s\to 0,
\\
{\hat{k}}(\alpha, \nu, {\mu}; a; s)\to 0,&\quad\quad s{\hat{k}}(\alpha, \nu, {\mu}; a; s)\to \infty, \quad &\text{ for} \quad\quad s\to \infty
\end{array}
\end{equation}
which are satisfied by ${\hat{k}}(\alpha, \nu, {\mu}; a; s) = s^{-1 + {\mu - \alpha\nu }} (s^{\alpha} + a)^{\nu}$ calculated in \cite[Eq. (2.5)]{TRPrabhakar71}. 
Weaker condition (proposed in \cite{ECdOliveira19,DZhao19} and satisfied in our case) is specified by $[s {\hat{k}}(\alpha, \nu, {\mu}; a; s)]^{-1}\to 0$ for $s\to\infty$. Then $k(\alpha, \nu, {\mu}; a, t)$ is called the {\em fading memory}, an old concept due to Ludwig Boltzmann, which, loosely speaking, says that a medium obeys it if changes happening in the past have less effect now than equivalent but more recent changes \cite{RSAnderssen02}. The fading memory condition is satisfied e.g. by completely monotone functions \cite{RSAnderssen02a}, the property which according to \cite[Theorem 2]{Kochubei2011} is guaranteed by conditions listed in Eq. \eqref{17_08_1}. 

We have announced earlier that the aim of our paper is to present solutions to the Volterra-type equation involving the memory kernel Eq. \eqref{5/04-1}.  General rules of the Laplace method say that taking $k(\alpha, \nu, \mu; a; t)$ whose Laplace transform exists and next applying the standard formalism  of the Laplace transform we can represent the solution of Eq. \eqref{12/03-2} as
\begin{equation}\label{20/03-1}
f(t) = \mathcal{L}^{-1}\left[\frac{\hat{k}(\alpha, \nu, \mu; a; s)}{s \hat{k}(\alpha, \nu, \mu; a; s) + B(\tau; \alpha, \nu, \mu)} f(0); t\right],
\end{equation}
where $\mathcal{L}^{-1}$ is the inverse Laplace transform. Depending on the ratio $B(\tau; \alpha, \nu, \mu)/[s\hat{k}(\alpha, \nu, \mu; a; s)]$ Eq. \eqref{20/03-1} can be considered two-fold. The first possibility is to extract $K(\alpha, \nu; a; s)$ from the nominator and denominator of Eq. \eqref{20/03-1} and use $(1+x)^{-1} = \sum_{r=0}^{\infty} (-x)^{r}$ for $x = B(\tau; \alpha, \nu, \mu)/[s\hat{k}(\alpha, \nu, \mu; a; s)]$. In such a case Eq. \eqref{20/03-1} reads
\begin{equation}\label{21/03-1}
f_{1}(\alpha, \nu, \mu; B, a; t) = \sum_{r=0}^{\infty} (-1)^{r} B^{\,r}(\tau; \alpha, \nu, \mu) \mathcal{L}^{-1}[s^{-1-r} \hat{k}^{-r}(\alpha, \nu, \mu; a; s); t] f(0) 
\end{equation}
satisfied for $|B(\tau; \alpha, \nu, \mu)/[s \hat{k}(\alpha, \nu, \mu; a; s)]| < 1$. This case was  discussed in \cite{ECdOliveira19,RGarra18,DZhao19}. To realize the second possibility we extract  $B(\tau; \alpha, \nu, \mu)$ and apply the series expansion of  $(1+x)^{-1}$  but this time for $x = \hat{k}(\alpha, \nu, \mu; a; s)/B(\tau; \alpha, \nu, \mu)$. Now, the solution to the Laplace transformed Eq. \eqref{20/03-1} can be represented as
\begin{equation}\label{21/03-2}
f_{2}(\alpha, \nu, \mu; B, a; t) = \sum_{r=0}^{\infty} (-1)^{r} B^{-1-r}(\tau; \alpha, \nu, \mu) \mathcal{L}^{-1}[s^{r} \hat{k}^{1+r}(\alpha, \nu, \mu; a; s); t] f(0),
\end{equation}
well defined for $|B(\tau; \alpha, \nu, \mu)/[s \hat{k}(\alpha, \nu, \mu; a; s)]| > 1$. This case, omitted in the papers \cite{RGarra18,DZhao19}, was discussed in \cite{Gorska19}.  In Secs. \ref{sec4} - \ref{sec7} we shall show how to sum up both series using methods of the umbral calculus and Laplace transform. 

\section{The three parameter Mittag-Leffler function $ \boldsymbol{E^{\,\nu}_{\alpha, 1 + d}(x)}$}\label{sec2}

The Prabhakar function $e^{\nu}_{\alpha, \mu}(\lambda, t)$ contains the three parameter Mittag-Leffler function $E^{\nu}_{\alpha, \mu}(x)$ whose series form is 
\begin{equation}\label{14/03-3}
E^{\nu}_{\alpha, \mu}(x) \okr \sum_{r=0}^{\infty}\frac{(\nu)_{r}\, x^{r}}{r!\, \Gamma(\alpha r + \mu)}, \quad \alpha, \mu, \nu > 0.
\end{equation}
The Pochhammer symbol (called also the raising factorial) is given by $(\nu)_{r} = \Gamma(\nu + r)/\Gamma(\nu) = \nu (\nu + 1) \ldots (\nu + r - 1)$. The three parameter Mittag-Leffler function for $\mu = \nu = 1$ tends to the one 
parameter or standard Mittag-Leffler function $E_{\alpha}(x) = E_{\alpha, 1}^{1}(x)$, whereas for $\nu = 1$ it is the two parameter Mittag-Leffler (Wiman) function $E_{\alpha, \mu}(x) = E_{\alpha, \mu}^{1}(x)$ (for a complete information on the Mittag-Leffler functions see\, {\em e.g.}, \cite{Mainardi-b-ML}). 

\subsection*{Umbral representations of $E^{\,\nu}_{\alpha, 1 + d}(x)$} 

In this Section we introduce the umbral representation of $E^{\,\nu}_{\alpha, 1 + d}(x)$ {which allow us to derive a set of handy formulae. \footnote{\, It must be mentioned that umbral methods usually do not provide us with adequate information on conditions under which they work and thus are treated as formal.}} Similarly to the classical references on the umbral calculus \cite{SRoman-b-84} we start our considerations introducing the shift operator $c_z$ defined as $c_{z}^{\,a}: h(z) \mapsto h(z + a)$ which satisfies $c^{\,a_{1}}_{z} c^{\,a_{2}}_{z} = c_{z}^{\,a_{1} + a_{2}}$. When we chose as $h(z)$ the function $[\Gamma(1 + z)]^{-1}$ then, by definition, $c_{z}^{\,a}\, [\Gamma(1+z)]^{-1}\big\vert_{z=b} = [\Gamma(1 + b + a)]^{-1}$. Thus we can define the umbral form of the three parameter Mittag-Leffler function as 
\begin{equation}\label{14/03-7}
E^{\nu}_{\alpha, 1 + d}(x) = c_{z}^{d} (1 - x c_{z}^{\alpha})^{-\nu} [\Gamma(1+z)]^{-1}\big\vert_{z = 0},
\end{equation}
where for our future applications we separate out $1$ in the parameter $\mu$ such that $\mu = 1 + d$, $d \geq -1$.

\subsection*{Properties of $E^{\,\nu}_{\alpha, 1 + d}(x)$} 

Below we present examples which show how the known formulae satisfied by $E^{\,\nu}_{\alpha, 1 + d}(x)$ can be derived with the help of calculations involving the umbral form Eq. \eqref{14/03-7} and using throughout the proofs tricky methods of the umbral calculus. 
{
\begin{proposition}
The Laplace transform of the Prabhakar function $e^{\nu}_{\alpha, 1+d}(a, x) = x^{d} E_{\alpha, 1 + d}^{\nu}(a x^{\alpha})$ (given in Refs. \cite{ Mainardi-b-ML, KGorska18a, TRPrabhakar71}) for $a \in \mathbb{R}$, $\RE(1 + d) > 0$, and $\RE(s) > 0$, $|s| > a^{1/\RE(\alpha)}$ reads 
\begin{equation}\label{16/03-1}
\mathcal{L}[e^{\nu}_{\alpha, 1+d}(a, x); s] = s^{\alpha\nu-1-d} (s^{\alpha} - a)^{-\nu}.
\end{equation}
\end{proposition}}
\noindent
{\sc {Proof of Eq. \eqref{16/03-1}.}} From Eq. \eqref{14/03-7} we have
\begin{align*}
\begin{split}
\mathcal{L}[e^{\nu}_{\alpha, 1+d}(a, x); s] & = \int_{0}^{\infty} \E^{- x s} x^{d} c_{z}^{d} (1 - a x^{\alpha} c_{z}^{\alpha})^{-\nu} [\Gamma(1+z)]^{-1}\big\vert_{z = 0} \D x \\
& = \sum_{r=0}^{\infty} \frac{a^r (\nu)_{r}}{r!}\, c_{z}^{d + \alpha r} [\Gamma(1+z)]^{-1}\big\vert_{z = 0} \int_{0}^{\infty} \E^{-x s} x^{d + \alpha r} \D x \\
& = s^{-1 - d} \sum_{r=0}^{\infty} \frac{(a s^{-\alpha})^r (\nu)_{r}}{r!}, 
\end{split}
\end{align*}
where in the first equality we have treated the umbral operator $c_{z}$ as a parameter and in the next step we have changed the order of summation and integration. The integral representation of the gamma function gives $s^{-1-d-\alpha r} \Gamma(\alpha r + 1 + d)$ and such obtained gamma function cancels with the gamma function which comes from $c_{z}^{d + \alpha r} [\Gamma(1+z)]^{-1}\big\vert_{z=0}$. Using once again the series expansion of $(s^{\alpha} + a)^{-\nu}$ {we complete} the proof. 
{
\begin{proposition}
The two parameters Mittag-Leffler function $E_{\alpha, 1 + d}(ax^{\alpha}) = E^{1}_{\alpha, 1 + d}(ax^{\alpha})$ with $ 0 < \RE(\alpha) < 1$, $\RE(d) > \RE(\alpha) - 1$, multiplied by $x^{d}$ is the eigenfunction of the $\alpha$-order fractional derivative in the Caputo sense:
\begin{equation}\label{30/04-0}
{^{\rm C}D^{\,\alpha}_{x}} [x^{\,d}E_{\alpha, 1 + d}(a x^{\alpha})] = a x^{\,d} E_{\alpha, 1 + d}(a x^{\alpha}), \qquad a\in\mathbb{R},
\end{equation}
where
\begin{equation*}
{^{\rm C}D^{\,\alpha}_{x}} h(x) = \frac{1}{\Gamma(1-\alpha)}\int_{0}^{t} (t - \xi)^{-\gamma} \frac{\D h(\xi)}{\D\xi} \D\xi  \qquad \text{for} \qquad \alpha\in(0, 1).
\end{equation*}
\end{proposition}}
\noindent
{\sc {Proof of Eq. \eqref{30/04-0}.}} Taking the definition of the fractional derivative in the Caputo sense, Eq. \eqref{14/03-7} and the series expansion of  $(1 - a x^{\alpha} c_{z}^{\alpha})^{-1}$ we get
\begin{align}\label{16/03-3}
\begin{split}
{^{\rm C}D^{\,\alpha}_{x}} [x^{\,d}E_{\alpha, 1 + d}(a x^{\alpha})] &= \int_{0}^{\infty} \frac{\frac{\D}{\D y}[y^{d}c_{z}^{d} (1-a y^{\alpha} c_{z}^{\alpha})^{-1}]}{(x-y)^{\alpha}}  [\Gamma(1+z)]^{-1}\big\vert_{z = 0} \frac{\D y}{\Gamma(1-\alpha)}  \\
& = \sum_{r=0}^{\infty} a^r c_{z}^{d + \alpha r}  [\Gamma(1+z)]^{-1}\big\vert_{z = 0}  \int_{0}^{\infty} \frac{\frac{\D}{\D y}y^{\,d + \alpha r}}{(x-y)^{\alpha}} \frac{\D y}{\Gamma(1-\alpha)}.
\end{split}
\end{align}
The integral in the second line of Eq. \eqref{16/03-3} is calculated using  the formula \cite[Eq. (2.2.4.8)]{APPrudnikov-v1}{, namely} $\int_{0}^{a} x^{\alpha-1} (a^{\mu} - x^{\mu})^{\beta-1} \D x = a^{\mu(\beta - 1) + \alpha} \mu^{-1} \Gamma(\alpha/\mu) \Gamma(\beta)/\Gamma(\alpha/\mu + \beta)${,} which completes the proof.  
{
\begin{proposition}
 For $\RE(\alpha) > 0$, $\RE(1+d-n) > 0$, $n\in\mathbb{N}$, and $a\in\mathbb{R}$ derivatives  of the Prabhakar function satisfy
\begin{equation}\label{16/03-4}
\frac{\D^{\,n}}{\D x^{n}} [x^{d} E^{\nu}_{\alpha, 1+d}(a x^{\alpha})] = x^{d - n} E^{\nu}_{\alpha, 1 + d - n}(a x^{\alpha}).
\end{equation}
\end{proposition}}

\smallskip
\noindent
{\sc {Proof of Eq. \eqref{16/03-4}.}} Expanding $(1 - a x^{\alpha} c_{z}^{\alpha})^{-\nu}$ in Eq. \eqref{14/03-7} as a series and substituting the result into the RHS  of Eq. \eqref{16/03-4} leads to
\begin{align*}
\begin{split}
\text{RHS of Eq. \eqref{16/03-4}} & = \sum_{r=0}^{\infty} \frac{a^r (\nu)_{r}}{r!} c_{z}^{d + \alpha r} [\Gamma(1+z)]^{-1}\big\vert_{z = 0} \, \frac{\D^{\;n}}{\D x^{\,n}} x^{d + \alpha r} \\
& = \sum_{r=0}^{\infty} \frac{a^r (\nu)_{r}}{r!} \frac{1}{\Gamma(1+d+\alpha r)} \frac{\Gamma(1 + d + \alpha r)}{\Gamma(1 + d - n + \alpha r)} x^{\alpha r + d -n}. 
\end{split}
\end{align*}
The above is the series representation of the three parameter Mittag-Leffler function multiplied by the power function $x^{d-n}$, \textit{i.e.} the LHS of Eq. \eqref{16/03-4}. 

\medskip
\begin{proposition}
The two parameter Mittag-Leffler functions satisfy the equalities (see formulae (10) and (13) of \cite{JWHanneken09})
\begin{equation}\label{13/04-2}
E_{-\alpha, 0}(x) = -x^{-1}E_{\alpha, \alpha}(x^{-1}) = - E_{\alpha, 0}(x^{-1}).
\end{equation}
\end{proposition}
{\sc{Proof of Eq. \eqref{13/04-2}.}} {The first equality in Eq. \eqref{13/04-2} stems directly from Eq. \eqref{14/03-7} applied to $E_{-\alpha, 0}(x)$. The second equality of Eq. \eqref{13/04-2} can be derived from the series representation of the two parameter Mittag-Leffler function in which we change the summation index $r = 0, 1, \ldots$ to $n = r+1$ where $n = 1, 2, \ldots$. Notice that we can do that because of $\lim_{r\to 0}y^{r}/\Gamma(\alpha r) = 0$. Hence, $E_{\alpha, \alpha}(y) = \sum_{n=1}^{\infty} y^{n-1}/\Gamma(\alpha n)= y^{-1} E_{\alpha, 0}(y)$.} 
{
\begin{proposition}
For $\alpha \in (0, 1)$, $\beta, \gamma > 0$, $a\in\mathbb{R}$, and $t > 0$ we have
\begin{equation}\label{11/02-10}
E_{\alpha, \beta}^{\gamma}(-a t^{\alpha}) = \frac{1}{\Gamma(\gamma)} \int_{0}^{\infty} \E^{-a u} t^{1-\beta} u^{\beta/\alpha - 1} g_{\alpha, \beta}^{\gamma}(u, t) \D u,
\end{equation}
where $g_{\alpha, \beta}^{\gamma}(u, t) = u^{-1/\alpha} g_{\alpha, \beta}^{\gamma}(t u^{-1/\alpha}) = u^{\gamma - \beta/\alpha} \mathcal{L}^{-1}[s^{\alpha\gamma - \beta} \E^{-u s^{\alpha}}; t]$.
\end{proposition}}
\noindent
{
{\sc {Proof of Eq. \eqref{11/02-10}.}} From Eq. \eqref{14/03-7} it appears that 
$E^{\,\gamma}_{\alpha, \beta}(-at^{\alpha}) = c_{z}^{\,\beta - \alpha\gamma -1} (c_{z}^{\, -\alpha} + at^{\alpha})^{-\gamma}\left[\Gamma(1+z)\right]^{-1}\vert_{z=0}$. Using the integral representation of $(c_{z}^{\, -\alpha} + at^{\alpha})^{-\gamma}$ we get
\begin{equation}\label{23/08-3}
E^{\,\gamma}_{\alpha, \beta}(-at^{\alpha}) = c_{z}^{\,\beta - \alpha\gamma -1}\int_{0}^{\infty}\D \xi\E^{-\xi\left(c_{z}^{\, -\alpha} + at^{\alpha}\right)}\frac{\xi^{\gamma -1}}{\Gamma(\gamma)}\, \left[\Gamma(1+z)\right]^{-1}\vert_{z=0}.
\end{equation}
The next step is to show that the umbral image $c_{z}^{\beta - \alpha\gamma -1} \exp(-\xi c_{z}^{-\alpha})[\Gamma(1+z)]^{-1}\vert_{z=0}$ is proportional to $g_{\alpha, \beta}^{\gamma}(\xi^{-1/\alpha})$ whose series form is presented in Appendix A: 
\begin{align*}
\begin{split}
c_{z}^{\beta - \alpha\gamma -1} \exp(-\xi c_{z}^{-\alpha})& [\Gamma(1+z)]^{-1}\vert_{z=0} = \sum_{r=0}^{\infty} \frac{(-\xi)^{r}}{r!} c_{z}^{\beta - \alpha\gamma-\alpha r - 1}[\Gamma(1+z)]^{-1}\vert_{z=0} \\
& = \xi^{(\beta-1)/\alpha - \gamma} \sum_{r=0}^{\infty} \frac{(-1)^{r}}{r! \Gamma(\beta - \alpha\gamma - \alpha r)} (\xi^{-1/\alpha})^{-1-\alpha r + \beta - \alpha\gamma}  \\
& = \xi^{(\beta-1)/\alpha - \gamma} g_{\alpha, \beta}^{\gamma}(\xi^{-1/\alpha}).
\end{split}
\end{align*}
Thus, Eq. \eqref{23/08-3} reads
\begin{equation*}
E^{\,\gamma}_{\alpha, \beta}(-at^{\alpha}) = \int_{0}^{\infty} \E^{-\xi a t^{\alpha}} \frac{\xi^{\beta/\alpha - (1+1/\alpha)}}{\Gamma(\gamma)} g_{\alpha, \beta}^{\gamma}(\xi^{-1/\alpha}) \D\xi.
\end{equation*}
Setting $\xi = t^{-\alpha} u$ we complete the proof.} 

\smallskip
\noindent
{
{\bf Remark 1.} In \cite{KGorska18a} it has  been derived the integral representation
\begin{equation}\label{23/08-1}
E^{\,\gamma}_{\alpha, \alpha\gamma}(-at^{\alpha})=\frac{1}{\Gamma(\gamma)}\int_{0}^{\infty}\E^{-au}t^{1-\alpha\gamma}{u^{\gamma -1}}\widetilde{\Phi}_{\alpha}(u, t)\D u,
\end{equation}
where $0 < \alpha < 1$, $\gamma > 0$, $a\in\mathbb{R}$, and $t > 0$. The function {$\widetilde{\Phi}_{\alpha}(u ,t) = u^{-1/\alpha} \widetilde{\Phi}_{\alpha}(t u^{-1/\alpha}) = \mathcal{L}^{-1}\left[\E^{-us^{\alpha}};t\right]$} denotes the one-sided L\'{e}vy distribution \cite{Pollard46}. This results from Eq. \eqref{11/02-10} for $\beta = \alpha\gamma$ for which $g_{\alpha, \alpha\gamma}^{\gamma}(u, t) = \widetilde{\Phi}_{\alpha}(u, t)$.}

\subsection*{The generalized hypergeometric form}

Rewriting \cite[Eq. (9)]{KGorska18a} we can express the three parameter Mittag-Leffler function for rational $0 < \alpha = l/k < 1$ as a finite sum of generalized hypergeometric functions
\begin{equation}\label{7/04-1}
E^{\,\nu}_{\frac{l}{k}, 1 + d}(x) = \sum_{j=0}^{k-1} \frac{x^{j}\, (\nu)_{j}}{j! \Gamma(1 + d + \frac{l}{k}j)} {_{1 + k}F_{l + k}}\left({1, \Delta(k, \nu +j) \atop \Delta(k, 1+j), \Delta(l, 1 + d + \frac{l}{k} j)}; \frac{x^{k}}{l^{l}} \right),
\end{equation}
where we adopt the symbol $\Delta(n, a)$ to denote the sequence $a/n, (a+1)/n, \ldots$, $(a+n-1)/n$ and to encode parameters of the generalized hypergeometric function. The upper (first) list of parameters is equal to $1$ and $\Delta(k, \nu + j)$, while the lower list, namely $\Delta(k, 1 + j)$ and $\Delta(l, 1 + d + l j/k)$, is the second lists of parameters. Note that the arguments of generalized hypergeometric functions in Eq. \eqref{7/04-1} are equal to $x^{k}/l^{l}$. 

\medskip
\noindent
Below we itemize a few examples of Eq. \eqref{7/04-1} which we will use in forthcoming parts of the paper: \\
\noindent
{(A)} Eq. \eqref {7/04-1} for $l/k = \nu = 1$ and $d = 1$ yields to 
\begin{equation*}
E^{1}_{1, 2}(x) = \frac{\exp(x) - 1}{x},
\end{equation*}
whereas for $l/k = 1$, $\nu = -1/2$, and $d = 1/2 + r$, $r = 0, 1, 2, \ldots$, it is equal to
\begin{equation}\label{26/04-2}
E^{\,-1/2}_{1, 3/2 + r}(x) = \frac{\exp(x)}{\Gamma(\frac{3}{2} + r)} {_{1}F_{1}}\left({2 + r \atop 3/2 + r}; -x\right). 
\end{equation}
In Eq. \eqref{26/04-2} we used the representation of the three parameter Mittag-Leffler function given by Eq. \eqref{7/04-1} and the Kummer relation for ${_{1}F_{1}}$.

\medskip
\noindent
{(B)} Eq. \eqref{7/04-1} for rational $\alpha = l/k$, $\nu = 1$, and $d = 0$ is the one parameter, or standard, Mittag-Leffler function:
\begin{equation}\label{7/04-2}
E_{l/k}(x) = \sum_{j=0}^{k-1} \frac{x^{j}}{\Gamma(1 + l j/k)} {_{1}F_{l}}\left({1 \atop \Delta(l, 1 + l j/k)}; \frac{x^{k}}{l^{l}}\right).
\end{equation}
As a particular case Eq. \eqref{7/04-2} reconstructs the well-known form of the one parameter Mittag-Leffler function for $l/k = 1/2$ \cite{Mainardi-b-ML}. Namely, from Eq. (7.11.2.19) of \cite{APPrudnikov-v3} we can express $E_{1/2}(x)$ as
\begin{equation}\nonumber
E_{1/2}(x) = \E^{x^{2}} [1 + {\rm erf}(x)],
\end{equation}
where ${\rm erf}(x)$ is the error function.

\medskip
\noindent
{(C)} In the case of the two parameter Mittag-Leffler function $E_{l/k, 0}(x)$, i.e. Eq. \eqref{7/04-1} written down for rational $\alpha = l/k$, $\nu = 1$, and $d = -1$, we get
\begin{equation}\label{10/4-5}
E_{l/k, 0}(x) = \sum_{j=1}^{k} \frac{x^{j}}{\Gamma(l j/k)} {_{1}F_{l}}\left({1 \atop \Delta(l, l j/k)}; \frac{x^{k}}{l^{l}}\right).
\end{equation}
Notice that the summation index $j$ is shifted and now it goes from 1 to $k$ (see the comment below Eq. \eqref{13/04-2}). Eq. \eqref{10/4-5} for $l = k = 1$  reads
\begin{equation*}
E_{1, 0}(x) = x\exp(x),
\end{equation*}
whereas for $l/k = 1/2$ it leads to  the combination of the power function and the one parameter Mittag-Leffler function $E_{1/2}(x)$:
\begin{equation*}
E_{1/2, 0}(x) = \frac{x}{\sqrt{\pi}} + x^{2} E_{1/2}(x). 
\end{equation*}

\section{The Mittag-Leffler polynomials $\boldsymbol{E^{\,-n}_{\alpha, 1 + d}(x)}$}\label{sec3}

For $\nu$ being a negative integer, i.e. $\nu = -n$, $n=1, 2, \ldots$, the three parameter Mittag-Leffler functions become polynomials called by us the Mittag-Leffler polynomials, namely
\begin{equation}\label{12/03-4}
E^{-n}_{\alpha, 1+ d}(x) \okr  \sum_{r=0}^{n} \binom{n}{r} \frac{(-x)^r}{\Gamma(\alpha r + 1 + d)}, \qquad \alpha > 0, \quad d \geq -1.
\end{equation} 
For natural $\alpha$, $\alpha = N\in\mathbb{N}$, they are related to the Konhauser polynomials $Z_{n}^{d}(x; N) =\Gamma(N n + d + 1) E^{-n}_{N, 1+d}(x^{N})/n!$ \cite{RGarra18, JDEKonhauser67, HMSrivastava82, TRPrabhakar71} which for $d > 0$ extend the associated Laguerre polynomials $L^{d}_{n}(x) = Z_{n}^{d}(x; 1)$ \cite{JDEKonhauser67}. The Mittag-Leffler polynomials $E^{-n}_{\alpha, 1+d}(x^{\alpha})$ for $d > 0$ belong also to the family of Laguerre-Wright polynomials $\Lambda^{(d, \alpha)}_{n}(y, x)$ \cite{DBabusi17} whose  generating functions are given in \cite{DBabusi17}.

\subsection*{Umbral representation of $E^{\,-n}_{\alpha, 1 + d}(x)$} 

The umbral representation \eqref{14/03-7} used for $\nu = -n$, $n=0, 1, 2, \ldots$ enables us to postulate the umbral image of the Mittag-Leffler polynomials Eq. \eqref{12/03-4} in the form
\begin{equation}\label{12/03-7a}
E^{-n}_{\alpha, 1 + d}(x) = c^{d}_{z} (1 - x c^{\alpha}_{z})^{n} \, [\Gamma(1+z)]^{-1}\big\vert_{z = 0}. 
\end{equation}
As seen from the previous paragraph the Mittag-Leffler polynomials $E^{-n}_{\alpha, 1 + d}(x^{\alpha})$ differ from the Konhauser polynomials ${Z}^{\,d}_{n}(x; \alpha)$  only by the normalization coefficient $n!/\Gamma(\alpha n + 1 + d)$ and $\alpha$ being real positive instead of natural. Thus, all algebraic formulae satisfied by the Konhauser polynomials $Z^{d}_{n}(x; N)$ and listed in \cite[Subsec. 3.II]{HMSrivastava82} as well as in \cite[Sec. II]{JDEKonhauser67} are also fulfilled by $E_{\alpha, \beta}^{-n}(x)$. 

\subsection*{Properties of $E^{\,-n}_{\alpha, 1 + d}(x)$} 

\begin{proposition}
{The derivative of the Mittag-Leffler polynomials satisfies}
\begin{equation}\label{13/03-1}
\frac{\D}{\D x} E^{\,-n}_{\alpha, 1 + d}(x^{\alpha}) = - \alpha\, x^{\alpha-1} E^{\,1-n}_{\alpha, 1 + d + \alpha}(x^{\alpha}), 
\end{equation}
from which one gets
\begin{equation}\label{13/03-2}
x \frac{\D}{\D x} E^{\,-n}_{\alpha, 1 + d}(x^{\alpha}) = n \alpha \left[E^{\,-n}_{\alpha, 1 + d}(x^{\alpha}) - E^{\,1-n}_{\alpha, 1 + d}(x^{\alpha})\right].
\end{equation}
\end{proposition}
\noindent
{\sc {Proof of Eq. \eqref{13/03-2}.}} Using the umbral representation Eq. \eqref{12/03-7a} of the Mittag-Leffler polynomials one has
\begin{align}\label{13/03-3}
\begin{split}
& x \frac{\D}{\D x} E^{\,-n}_{\alpha, 1 + d}(x^{\alpha}) = n \alpha\,c_{z}^{\,d}\, \frac{(1 - x^{\alpha} c_z^{\alpha}) - 1}{1 - x^{\alpha} c_{z}^{\alpha}} (1 - x^{\alpha}c_{z}^{\alpha})^{n} [\Gamma(1+z)]^{-1}\big\vert_{z=0} \\
& \qquad\quad = n \alpha \left\{ c_{z}^{\,d}\, (1 - x^{\alpha}c_{z}^{\alpha})^{n} - c_{z}^{\,d}\, (1 - x^{\alpha}c_{z}^{\alpha})^{n-1}\right\}[\Gamma(1+z)]^{-1}\big\vert_{z=0}
\end{split}
\end{align}
which gives Eq. \eqref{13/03-2}. 

\medskip
\noindent
{\bf{Remark 2.}} Eqs. \eqref{13/03-1} and \eqref{13/03-2} lead to similar but not the same formulae as those obeyed by the standard Konhauser polynomials being exhibited in Eqs. (3.10) and (3.12) of \cite{HMSrivastava82}. The difference is caused by the fact that now we are dealing with the real positive parameter $\alpha$. Additionally, from the first equality in Eq. \eqref{13/03-3} we derive 
\begin{equation}\nonumber
x^{\alpha} E^{\,-n}_{\alpha, 1 + d+\alpha}(x^{\alpha}) + E^{\,-(n+1)}_{\alpha, 1 + d}(x^{\alpha}) = E^{\,-n}_{\alpha, 1 + d}(x^{\alpha}).
\end{equation}
{
\begin{proposition}
If $a, b > 0$ then
\begin{equation}\label{14/03-4}
\int_{0}^{\infty}\!\! x^{d + m}\! \E^{-a x}\! E^{\,-n}_{\alpha, 1 + d}[(b x)^{\alpha}]\! \D x =  a^{-1-d-m} \sum_{r=0}^{n}\!\! \binom{n}{r} \!\left[-\!\left(\frac{b}{a}\right)^{\!\! \alpha}\,\right]^{r} \frac{\Gamma(1+d+m+\alpha r)}{\Gamma(1+d+\alpha r)},
\end{equation}
which vanishes for $a = b$ and $m=0,1$. 
\end{proposition}
}
\noindent
{\sc {Proof of Eq. \eqref{14/03-4}.}} Substitution of the series representation of Eq. \eqref{12/03-4} into the LHS of Eq. \eqref{14/03-4} leads to
\begin{equation}\nonumber
\text{RHS of Eq. \eqref{14/03-4}} =  \sum_{r=0}^{n} \binom{n}{r} (-b^{\alpha})^{r} c_{z}^{d+\alpha r }[\Gamma(1+z)]^{-1}\big\vert_{z=0} \int_{0}^{\infty}\!\! \E^{-a x} x^{d + m + \alpha r} \D x, 
\end{equation}
in which we use the umbral form of $[\Gamma(\alpha r + 1 + d)]^{-1}$, i.e. the action of the umbral operator $c_{z}^{d+\alpha r }$ on $[\Gamma(1+z)]^{-1}$ at $z=0$. Applying the integral representation of the gamma function, we obtain the RHS of Eq. \eqref{14/03-4}. 

\noindent
Vanishing of the integral in Eq. \eqref{14/03-4}, if $a=b$, is seen from \cite[Eq. (2)]{JDEKonhauser67} rewritten for the Mittag-Leffler polynomials. For $m=0$ and $m=1$ it is easy to show - for $m=0$ and from Eq. \eqref{14/03-4} we get
\begin{equation}\nonumber
\int_{0}^{\infty} x^{d} \E^{-a x} E^{\,-n}_{\alpha, 1 + d}[(b x)^{\alpha}] \D x = a^{-1-d-\alpha n} (a^{\alpha} - b^{\alpha})^{n},
\end{equation}
whereas for $m=1$ we have
\begin{equation}\nonumber
\int_{0}^{\infty} x^{d + 1} \E^{-a x} E^{\,-n}_{\alpha, 1 + d}[(b x)^{\alpha}] \D x = a^{-1-d-\alpha n} (a^{\alpha} - b^{\alpha})^{n-1} [(1+d) a^{\alpha} - (1+d+\alpha n) b^{\alpha}].
\end{equation}
RHS's of both these expressions  vanish for $a=b$ which finishes the proof. 

\subsection*{The generalized hypergeometric form}

We complete the information about the Mittag-Leffler polynomials by expressing them as the finite sum of the generalized hypergeometric functions. From the definition of the three parameter Mittag-Leffler function given via Eq. \eqref{7/04-1} for the rational $\alpha = l/k$ in which we apply $(-n)_{r} = (-1)^{r}/(n)_{-r}$ one finds
\begin{equation}\label{12/03-7}
E^{-n}_{\frac{l}{k}, 1 + d}(x) = \sum_{j=0}^{M}\!\! \binom{n}{j} \frac{(-x)^j}{\Gamma(1 + d + \frac{l}{k} j)} {_{1+k}F_{l+k}}\left({1, \Delta(k, -n +j) \atop \Delta(k, 1+j), \Delta(l, 1+d + \frac{l}{k} j)}; \frac{x^{k}}{l^{l}}\right)\!,
\end{equation}
where $M$ is the smallest number from $n$ and $k-1$. For integer 
$\alpha=l/k$, where $l = N$ and $k = 1$, Eq. \eqref{12/03-7} tends to \cite[Eq. (3.2)]{HMSrivastava82}; this stems from the hypergeometric representation of the Konhauser polynomials.

\section{The integral kernel $\boldsymbol{e^{-1}_{\alpha, 1- \alpha}(-a, t)}$}\label{sec4}

As the first example of solving Eqs. \eqref{21/03-1} and \eqref{21/03-2} we will consider the model governed by Eq. \eqref{5/04-1} for $\nu = 1$ {and $\mu = \alpha$}. In this case, the integral kernel $k(\alpha, 1, {\alpha}; a; t)$ is equal to the Prabhakar function $e^{-1}_{\alpha, 1-\alpha}(-a, t)$ with $\alpha\in(0, 1)$ \cite{RGarra18, RGarrappa17, DZhao19a, DZhao19}.  The series in Eq. \eqref{21/03-1} reads as \cite[Eq. (10)]{DZhao19} and/or as \cite[Eq. (16)]{RGarra18}, namely
\begin{equation}\label{14/03-2}
f_{1}(\alpha, 1, {\alpha}; B, a; t) = \sum_{r=0}^{\infty} (-B)^{r}  e^{r}_{\alpha, 1 + \alpha r}(-a, t)f(0),  
\end{equation}
and the series in Eq. \eqref{21/03-2} is
\begin{equation}\label{12/03-3}
f_{2}(\alpha, 1, {\alpha}; B, a; t) = - \sum_{r=0}^{\infty} (-B)^{-1-r} e^{-1-r}_{\alpha, 1 - \alpha(1+r)}(-a, t) f(0), 
\end{equation}
where $B = B(\tau; \alpha, 1, \alpha)$. Summing up the series \eqref{14/03-2} and \eqref{12/03-3} will be done two-fold - applying the umbral images of Secs.~\ref{sec2} and \ref{sec3} or using the Laplace transform method. 

\subsection*{The umbral calculus method}

The Prabhakar function {$e_{\alpha, 1 + \alpha r}^{r}(-a, t)$} in Eq. \eqref{14/03-2} is the product of the power $t^{\alpha r}$ and the three parameter Mittag-Leffler functions $E^{\,r}_{\alpha, 1 + \alpha r}(-a t^{\alpha})$ whose umbral representation is given through Eq. \eqref{14/03-7}. Using the latter we can sum up the series in Eq. \eqref{14/03-2} and get
\begin{equation*}%\label{16/03-8}
f_{1}(\alpha, 1, {\alpha}; B, a; t) = \frac{1+ a t^{\alpha}c_{z}^{\alpha}}{1 + (B + a) t^{\alpha} c_{z}^{\alpha}} [\Gamma(1+z)]^{-1}\big\vert_{z=0}\, f(0),
\end{equation*}
which after applying twice the umbral form of the three parameter Mittag-Leffler function leads to
\begin{align}\label{16/03-10}
\begin{split}
f_{1}(\alpha, 1, {\alpha}; B, a; t) & = E_{\alpha}[-(B+a) t^{\alpha}] f(0) +a t^{\alpha} E_{\alpha, 1 + \alpha} [-(B+a) t^{\alpha}]\, f(0) \\
& = \frac{B}{B +a} E_{\alpha}[-(B+a) t^{\alpha}] f(0) + \frac{a}{B +a}\, f(0)\\
& = \frac{B}{B +a} \left[E_{\alpha}[-(B+a) t^{\alpha}]-1\right] f(0) + f(0) . 
\end{split}
\end{align}
The middle line in Eq. \eqref{16/03-10} comes from the upper one by employing the formula $E_{\alpha, 1 + \alpha}(\sigma) = \sigma^{-1} E_{\alpha}(\sigma) - \sigma^{-1}$ immediately derived from the series representation of $E_{\alpha, 1 + \alpha}(\sigma)$ (see Eq. \eqref{14/03-3} for $\nu =1$ and $\mu = 1 + \alpha$). Repeating the calculations for the series in Eq. \eqref{12/03-3} we obtain the same result for $f_{2}(\alpha, 1, {\alpha}; B, a; t)$.

\subsection*{The Laplace transform method}\label{ss2}

The Laplace transform of $e^{-1}_{\alpha, 1 -\alpha}(-a, t)$, calculated with the help of Eq. \eqref{16/03-1},  gives ${\hat{k}}(\alpha, 1, {\alpha}; a; s) = s^{-1}(s^{\alpha} + a)$. Thus, Eq. \eqref{20/03-1} yields to 
\begin{equation}\nonumber
f(\alpha, 1, {\alpha}; B, a; t) = \mathcal{L}^{-1}\left\{\frac{s^{\alpha-1} + a s^{-1}}{s^{\alpha} + (B + a)}; t\right\} f(0).
\end{equation}
Using once again Eq.  \eqref{16/03-1} we express $f(\alpha, 1, {\alpha}; B, a; t)$ as the difference of one and two parameter Mittag-Leffler functions given in the first line of  Eq. \eqref{16/03-10}.

The reader should not be surprised by this result because the definitions \eqref{5/04-1} and \eqref{12/03-4} say that the fading memory kernel is equal to $e^{-1}_{\alpha, 1-\alpha}(-a, t) = [t^{\alpha}\Gamma(1-\alpha)]^{-1} + a$. Such a substitution into the adjusted Eq. \eqref{12/03-2} gives
\begin{equation}\label{9/04-1}
{^{\rm C}D^{\alpha}_{x}} f(\alpha, 1, {\alpha}; B, a; t) = - (B + a) f(\alpha, 1, {\alpha}; B, a; t) + a f(0),
\end{equation}
which solution is Eq. \eqref{16/03-10} as has been noticed in \cite{KGorska12a, IPodlubny99, BJWest10}. Eq. \eqref{9/04-1} generalizes the Cole-Cole relaxation equation adding into it a non-zero inhomogenous  term which may be interpreted as an external ``constant force term'' $a f(0)$. Obviously, for $a = 0$ (identified with $\varepsilon(\infty)=0$ we get the standard Cole-Cole relaxation governed by the fractional equation with the Caputo derivative. 

\subsection*{A remark on asymptotics of $f(\alpha, 1, {\alpha}; B, a; t)$}

\begin{figure}[!h]
\begin{center}
\includegraphics[scale=0.5]{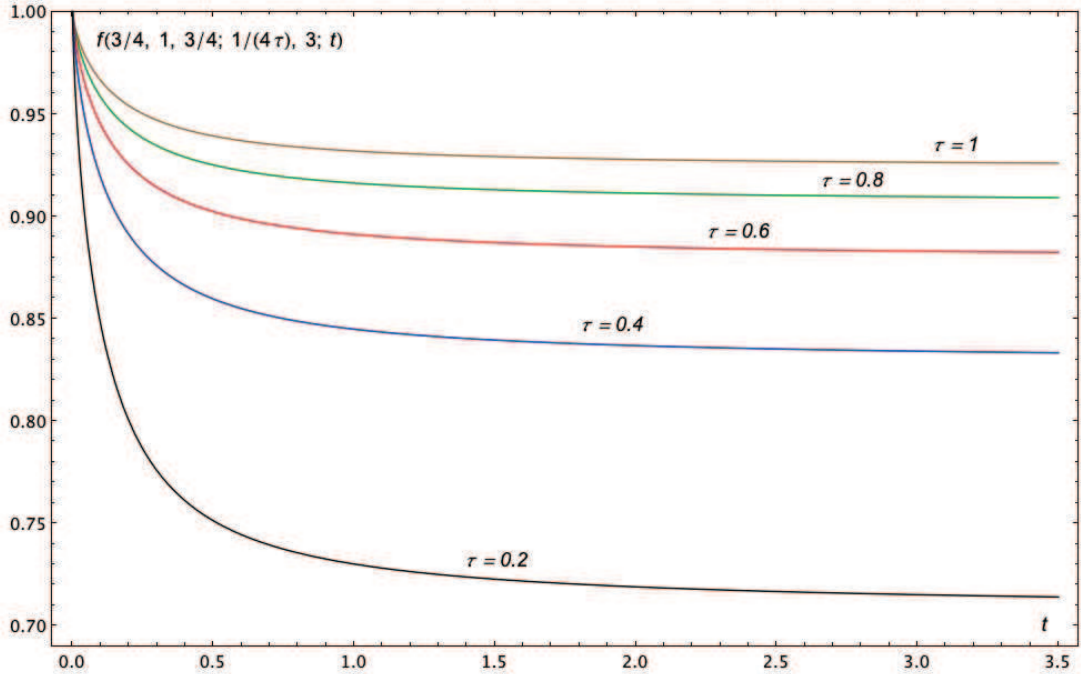}
\end{center}
\caption{\label{fig3} The solution given by Eq. \eqref{16/03-10} for $f(0) = 1$, $\alpha = 3/4$, $B = (4\tau)^{-1}$, $a = 3$, and $\tau = 0.2$ (the black curve), $\tau = 0.4$ (the blue curve), $\tau = 0.6$ (the red curve), $\tau = 0.8$ (the green curve), and $\tau = 1$ (the brown curve). The one parameter Mittag-Leffler function for $\alpha = 3/4$ is expressed by Eq. \eqref{7/04-2}.}
\end{figure}

The value of $f(\alpha, 1, {\alpha}; B, a; t)$ at $t=0$ is 1. That can be deducted from Eq. \eqref{16/03-10} and the fact that the one parameter Mittag-Leffler function equals 1 at zero. It is also seen from the asymptotics of $E_{\alpha}(\sigma)$. Due to the range of time we have two different asymptotic behaviors \cite{RGarrappa16}:
\begin{equation}\nonumber
E_{\alpha}(- a t^{\alpha}) \sim \left\{\begin{array}{c c} 1 - A t^{\alpha}/\Gamma(1+\alpha), \qquad  & \text{for} \quad A^{1/\alpha} t \ll 1 \\ (A t^{\alpha})^{-1}/\Gamma(1-\alpha), \qquad  & \text{for} \quad A^{1/\alpha} t \gg 1\end{array}\right. ,
\end{equation} 
where $A = B + a$. The first one, valid for $A^{1/\alpha} t \ll 1$, reconstructs the value of one parameter  Mittag-Leffler function at zero and from the second one it stems that
\begin{equation}\nonumber
\lim_{t\to\infty} f(\alpha, 1, {\alpha}; B, a; t) = \frac{a}{B + a} f(0) 
\end{equation}
Asymptotic behavior of  $f(\alpha, 1, {\alpha}; B, a; t)$ is illustrated in Fig. \ref{fig3} where we have plotted Eq. \eqref{16/03-10} for $\alpha = 3/4$, $B=(1-\alpha)/\tau$ and  $\tau = 0.2, 0.4, 0.6, 0.8, 1$. If $a\neq 0$ then for $t\to \infty$ the response function $f(\alpha, 1, {\alpha}; B, a; t)$ becomes almost constant but does not vanish which means existence of some residual polarization induced by a constant external field acting on a system.

\section{The integral kernel $\boldsymbol{e^{-\nu}_{\alpha, 1 - \alpha\nu}}(-a, t)$}\label{sec6}

The {next} example undergoing our study is Eq. \eqref{12/03-2} with the memory kernel $k(\alpha, \nu, {\alpha\nu}; a; t) = e^{-\nu}_{\alpha, 1 - \alpha\nu}(-a, t)$, $\alpha, \nu \in (0, 1)$ which the Laplace transform is ${\hat{k}}(\alpha, \nu, {\alpha\nu}; a; s) = s^{-1}(s^{\alpha} + a)^{\nu}$. Like it was done previously we shall solve Eqs. \eqref{21/03-1} and \eqref{21/03-2} using the umbral calculus and the Laplace transform method.  For the Laplace transformed memory kernel ${\hat{k}}(\alpha, \nu, {\alpha\nu}; a; s)$ Eqs. \eqref{21/03-1} and \eqref{21/03-2} read
\begin{equation}\label{6/05-1}
f_{1}(\alpha, \nu, {\alpha\nu}; B, a; t) = \sum_{r=0}^{\infty} (-B)^{r} e^{\nu r}_{\alpha, 1 + \alpha\nu r}(-a, t) f(0)
\end{equation}
and 
\begin{equation}\label{6/05-2}
f_{2}(\alpha, \nu, {\alpha\nu}; B, a; t) = - \sum_{r=0}^{\infty} (-B)^{-1-r} e^{-\nu (1+r)}_{\alpha, 1 - \alpha\nu (1+r)}(-a, t) f(0),
\end{equation}
where $B = B(\tau; \alpha, \nu, \alpha\nu)$.

\subsection*{The umbral method}
Let us begin with differentiating \eqref{6/05-1} with respect to time:
\begin{align*}
\begin{split}
\frac{\D}{\D t}f_{1}(\alpha, \nu, {\alpha\nu}; B, a; t) &= \sum_{r=0}^{\infty} (-B)^{r} \left[\frac{\D}{\D t}e^{\, \nu r}_{\alpha, 1+\alpha\nu r}(-a,t)\right] f(0) \\ &=\sum_{r=0}^{\infty} (-B)^{r} t^{\alpha\nu r -1}E^{\, \nu r}_{\alpha, \alpha\nu r}(-at^{\alpha})] f(0).
\end{split}
\end{align*}
Substituting next the integral representation of $E^{\, \nu r}_{\alpha, \alpha\nu r}(-at^{\alpha})$ given by Eq. \eqref{23/08-1} into the formula above we get
\begin{align}\label{23/08-8}
\frac{\D}{\D t}f_{1}(\alpha, \nu, {\alpha\nu}; B, a; t) = \int_{0}^{\infty}\E^{-au} \sum_{r=0}^{\infty} \frac{(- B u^{\nu})^{r} }{\Gamma(\nu r)}u^{-1}\widetilde{\Phi}_{\alpha}(u,t) \D u f(0).
\end{align}
with $\widetilde{\Phi}_{\alpha}(u,t)={\cal{L}}^{-1}[\exp{-us^{\alpha}};t]$ being the one-sided L\'{e}vy distribution. The sum in the integrand multiplied by $u^{-1}$  is $e_{\nu,0}^{\, 1}(-B,u)$ so introducing the function ${\Phi}_{\alpha}(u, t) = \int_{0}^{t}\widetilde{\Phi}_{\alpha}(u,t) \D t$ we rewrite down Eq. \eqref{23/08-8} as
\begin{equation}\label{23/08-8a}
f_{1}(\alpha, \nu, {\alpha\nu}; B, a; t) = f(0) + \int_{0}^{\infty}\E^{-au} e_{\nu,0}^{\, 1}(-B,u){\Phi}_{\alpha}(u,t) \D u f(0).
\end{equation}
Calculations performed for $f_{2}(\alpha, \nu, {\alpha\nu}; B, a; t)$ give the same result.  

\subsection*{The Laplace transform method}
Summing up the series either for $f_{1}(\alpha, \nu, {\alpha\nu}; B, a; t)$ or for $f_{2}(\alpha, \nu, {\alpha\nu}; B, a; t)$ without the help of Eq. \eqref{23/08-1} and using only the Laplace transform methods  requires rather laborious calculations. If done then their final result reads
\begin{multline}\label{6/05-4}
f(\alpha, \nu, {\alpha\nu}; B, a; t) = f_{1}(\alpha, \nu, {\alpha\nu}; B, a; t) = f_{2}(\alpha, \nu, {\alpha\nu}; B, a; t) \\ = f(0) + \int_{0}^{\infty} \E^{-a\xi} e^{1}_{\nu, 0}(-B, \xi) \varPhi_{\alpha}(\xi, t) \D\xi\, f(0),
\end{multline}
where the function $\varPhi_{\alpha}(\xi, t)$ is equal to $\mathcal{L}^{-1}[s^{-1} \E^{-\xi s^{\alpha}}; t]$. Obviously it means that its time derivative equals  to the one-sided L\'{e}vy stable distribution $\widetilde\Phi_{\alpha}(\xi, t)$ and we recover the Eq. \eqref{23/08-8}. Note also that $\varPhi_{\alpha}(\xi, t)$ for $\alpha = 1$ reduces to the Heaviside step function. 

\subsection*{A few remarks on the function $\varPhi_{\alpha}(\xi, t)$}

For rational $\alpha = l/k$ the function $\varPhi_{l/k}(\xi, t)$ can be expressed by  the Meijer $G$ function: 
\begin{equation}\label{6/05-5}
\varPhi_{l/k}(\xi, t) = \frac{\sqrt{k/l}}{(2\pi)^{(k-l)/2}} G^{k, 0}_{l, k}\left(\!\big(\ulamek{\xi}{k}\big)^{\!k} \big(\ulamek{l}{t}\big)^{\! l}\Big\vert {\Delta(l, 1) \atop \Delta(k, 0)}\right) 
\end{equation}
where the sequence $\Delta(n, a)$ is defined below Eq. \eqref{7/04-1}. Technically burdensome derivation of Eqs. \eqref{6/05-5} is shifted to Appendix A but its result is worthy to be quoted {\sl in extenso} because provides us with a tool which may be directly implemented in calculations using the standard algebra computer systems. Eq. \eqref{6/05-5} appears useful also for analytical calculations. 

\section{{The integral kernel $e_{\alpha, 1 -\mu}^{-\nu}(-a, t)$}}\label{sec7}

{
The example discussed in the previous section involves the kernel $e_{\alpha, 1 - \alpha\nu}^{\nu}(-a, t)$ depending on two parameters $\alpha, \nu \in (0, 1)$. We generalize it introducing the kernel $k(\alpha, \nu, \mu; a; t) = e^{-\nu}_{\alpha, 1 - \mu}(-a, t)$ with three independent parameters $\alpha, \nu, \mu\in (0, 1)$. Hence, the solutions of Eqs. \eqref{21/03-1} and \eqref{21/03-2} read
\begin{equation}\label{11/02-20}
f_{1}(\alpha, \nu, \mu; B, a; t) = \sum_{r=0}^{\infty} (-B)^{r} e_{\alpha, 1 + \mu r}^{\nu r}(-a, t) f(0)
\end{equation}
and
\begin{equation}\label{11/02-21}
f_{2}(\alpha, \nu, \mu; B, a; t) = -\sum_{r=0}^{\infty} (-B)^{-1-r} e_{\alpha, 1 - \mu (1 + r)}^{-\nu (1 + r)}(-a, t) f(0).
\end{equation}
To sum up these series we will use the method similar to that adopted for Eqs. \eqref{6/05-1}  and \eqref{6/05-2} but this time we apply Eq. \eqref{11/02-10} instead of its special case Eq. \eqref{23/08-1}. Derivation of $f_{1}(\alpha, \nu, \mu; B, a; t)$ with respect to time, with Eq. \eqref{16/03-4} employed, leads to
\begin{equation*}%\label{11/02-22}
\frac{\D}{\D t} f_{1}(\alpha, \nu, \mu; B, a; t) = \sum_{r=0}^{\infty} (-B)^{r} t^{\mu r} E_{\alpha, \mu r}^{\nu r}(-a t^{\alpha}) f(0).
\end{equation*}
Thereafter, we use the integral representation of the three parameter Mittag-Leffler function given through Eq. \eqref{11/02-10} with $u^{\beta/\alpha - \gamma} g_{\alpha, \beta}^{\gamma}(u, t)$ expressed as the inverse Laplace transform of $s^{\alpha\gamma - \beta} \exp(-u s^{\alpha})$. That implies
\begin{align}\label{11/02-23}
\begin{split}
\frac{\D}{\D t} f_{1}(\alpha, \nu, \mu; B, a; t) & = \int_{0}^{\infty} \E^{-a u} u^{-1} \sum_{r=0}^{\infty} \frac{(-B u^{\nu})^{r}}{\Gamma(\nu r)} \mathcal{L}^{-1}[s^{(\alpha\nu - \mu)r} \E^{- u s^{\alpha}}; t] \D u\\
& = \int_{0}^{\infty} \E^{-a u} \mathcal{L}^{-1}\left[u^{-1} \sum_{r=0}^{\infty}\frac{[-B s^{(\alpha\nu - \mu)} u^{\nu}]^{r}}{\Gamma(\nu r)} \E^{-u s^{\alpha}}; t\right] \D u\\
& = \int_{0}^{\infty} \E^{-a u} \mathcal{L}^{-1}[e_{\nu, 0}^{1}(- B s^{\alpha\nu - \mu}, u) \E^{-u s^{\alpha}}; t]\D u.
\end{split}
\end{align} 
The inverse Laplace transform in the last integrand above is unknown. If compared with Eqs. \eqref{23/08-8} and \eqref{23/08-8a} its calculation is obstructed (if $\alpha\nu - \mu\neq 0$) by the dependence of $e_{\nu, 0}^{1}(- B s^{\alpha\nu - \mu}, u)$ on the Laplace variable $s$.  Nevertheless it is not difficult to show that Eqs. \eqref{11/02-20} and \eqref{11/02-21} lead to the same result: calculating $\D f_{2}(\alpha, \nu, \mu; B, a; t)/\D t$ with the relation $E_{-\nu, 0}(x) = - E_{\alpha, 0}(x^{-1})$ used, then integrating  over the time and applying the initial condition $f(t=0) = f(0)$ we arrive at 
\begin{equation*}
f_{1}(\alpha, \nu, \mu; B, a; t) = f_{2}(\alpha, \nu, \mu; B, a; t) \equiv f(\alpha, \nu, \mu; B, a; t). 
\end{equation*}
Observe that Eq. \eqref{11/02-23} for $\mu = \alpha\nu$ leads to Eq. \eqref{6/05-4}. Similarly, the integral form of $E_{\,\alpha, \beta}^{\gamma}(-a t^{\alpha})$ for $\beta = \alpha\gamma$ reduces to Eq. \eqref{23/08-1}. }

\section{Discussion and conclusion}\label{sec8}

\subsection*{{The spectral function of non-Debye relaxation processes}}

{
The relation between the spectral function $\hat{\phi}(\alpha, \nu, \mu; \I\!\omega)$ and the relaxation function $f(\alpha, \nu, \mu; B, a; t)$ is $\hat{\phi}(\alpha, \nu, \mu; a; \I\!\omega) = \mathcal{L}[-\D f(\alpha, \nu, \mu; B, a; t)/\D t; \I\!\omega]$. Thus, using the properties of the Laplace transform, we have $f(\alpha, \nu; B, a; t) = \mathcal{L}^{-1}\{[1 - \hat{\phi}(\alpha, \nu, \mu; a; \I\!\omega)]/(\I\!\omega); t]$ \cite{KGorska18a}. Inserting Eq. \eqref{20/03-1} we get
\begin{equation}\label{9/02-3}
\hat{\phi}(\alpha, \nu, \mu; a; \I\!\omega) = \left[1 + \frac{\I\!\omega \hat{k}(\alpha, \nu, \mu; a; \I\!\omega)}{B(\tau; \alpha, \nu, \mu)}\right]^{-1},
\end{equation}
where we take $f(0) = 1$. Now we calculate the spectral function $\hat{\phi}(\alpha, \nu, \mu; a; \I\!\omega)$ involving the kernel $k(\alpha, \nu, \mu; a; t)$, given via Eq. \eqref{5/04-1}, for which $\hat{k}(\alpha, \nu, \mu; a; s) = s^{-1 + \mu - \alpha\nu} (s^{\alpha} + a)^{\nu}$. From Eq. \eqref{9/02-3} such kernel yields to 
\begin{equation*}%\label{12/02-2}
\hat{\phi}(\alpha, \nu, \mu; a; \I\!\omega) = \frac{B}{B + (\I\!\omega)^{\mu- \alpha\nu} [a + (\I\!\omega)^{\alpha}]^{\nu}}
\end{equation*}
with $B = B(\tau; \alpha, \nu, \mu)$. The spectral function, given as the ratio of the permittivities $[\tilde{\varepsilon}(\I\!\omega) - \varepsilon_{\infty}]/(\varepsilon_{0} - \varepsilon_{\infty})$ where $\varepsilon_{0}$ and $\varepsilon_{\infty}$ are permittivities in low and high frequencies, must equal 1 for $\omega = 0$.  The only possibility to satisfy this condition is to put $\mu=\alpha\nu$, $a=0$ and, because of dimensional reasons, $B = \tau^{-\mu}$. This leads to the spectral function appropriate for the Cole-Cole model 
\begin{equation}\label{10/02-11}
\hat{\phi}(\alpha, \nu, \mu; 0; \I\!\omega) = [1 + (\I\!\omega\tau)^{\mu}]^{-1}.
\end{equation}
The choice $B = \tau^{-\mu}$ guarantees the correct dimension $[\textrm{time}^{-1}]$ of the response function $\phi(\alpha, \nu, \mu; 0; t) = B t^{\mu - 1} E_{\mu, \mu}(- B t^{\mu})$. }

Discussing this part of our results we conclude that among the standard non-Debye relaxations described by the spectral function $\hat{\varPhi}(\I\!\omega) = [1 + (\I\!\omega\tau)^{\beta}]^{-\lambda}$ where $\beta, \lambda \in (0, 1]$ only the Cole-Cole case, i.e. $\lambda = 1$ and $\beta \in (0, 1)$, is reconstructed in the scheme based on the Prabhakar functions. For $a=0$ the Prabhakar kernel becomes power-like and Eq.\eqref{12/03-2} takes on the form of the fractional differential equation with the Caputo derivative. 

\subsection*{Jonscher's universal law}

{
The law describing the high and low frequency asymptotics of permittivity $\hat{\varepsilon}(\I\!\omega) = \hat{\phi}(\I\!\omega)(\varepsilon_{0} - \varepsilon_{\infty}) + \varepsilon_{\infty}$ was discovered and advocated by A. K. Jonscher and his collaborators on the basis of comprehensive analysis of experimental results \cite{AJonsher92}. According to this law properties of the vast majority of dielectric materials are characterized by parameters $m$ and $1-n$,  both from the range of $(0, 1]$, and this feature is independent from structural details of examined dielectrics. Jonscher's universal law reads
\begin{equation*}%\label{12/02-10}
\hat{\varepsilon}(\I\!\omega) \propto (\I\!\omega\tau)^{n-1} \quad \text{for} \quad \tau\omega \gg 1  \qquad \text{and} \qquad \Delta\hat{\varepsilon}(\I\!\omega) \propto (\I\!\omega\tau)^{m} \quad \text{for} \quad \tau\omega \ll 1,
\end{equation*}
where $\Delta\hat{\varepsilon}(\I\!\omega) = \varepsilon_{0} - {\hat{\varepsilon}}(\I\!\omega)$. For the majority of observed relaxation processes $m \geq 1 - n$; they are called typical and include the Cole-Cole ($m = 1 - n$), Cole-Davidson ($m=1 > 1-n$), and Havriliak-Negami ($m > 1-n$) models. If $m < 1-n$ then we deal with atypical processes described by \textit{e.g.} the Jurlewicz - Weron - Stanislavsky (JWS)  model \cite{RGarrappa16, AJurlewicz10, AStanislavsky10, AStanislavsky17, AStanislavsky19} - for a convenient illustration see Fig. (1) of \cite{AStanislavsky17}. Rewriting Eq. \eqref{10/02-11} in the form of \cite[Eq. (9)]{AStanislavsky18} and using the asymptotics given below Eq. (9)  there  we obtain that $n-1 = -\mu$ and $m = \mu - \alpha\nu$ with $\mu\in(0, 1]$ and $\alpha\nu \in[\mu-1, \mu)$. We remark that for $\mu = \alpha\nu$ and $a\neq 0$ the exponent $m = 0$ what is forbidden by the Jonscher's universal law but for $\mu = \alpha\nu$ and $a = 0$ we have the Cole-Cole model \eqref{10/02-11} for which Jonscher's universal law reads}
\begin{equation*}%\label{10/02-15}
{\hat{\varepsilon}(\I\!\omega) \propto (\I\!\omega\tau)^{-\mu} \quad \text{for} \quad \tau\omega \gg 1 \qquad \text{and} \qquad \Delta\hat{\varepsilon}(\I\!\omega)  \propto (\I\!\omega\tau)^{\mu} \quad \text{for} \quad \tau\omega \ll 1.}
\end{equation*}

\subsection*{{Link between Eqs. \eqref{9/02-1} and \eqref{12/03-2}}}

{
The stochastic nature of the spectral function $\hat{\phi}(\alpha, \nu, \mu; a; \I\!\omega)$ enables one to write 
\begin{equation}\label{12/02-1}
\hat{\phi}(s) = [1 + \Psi(s)/B]^{-1}, \qquad s\in\mathbb{C},
\end{equation}
where $B = B(\tau; \alpha, \nu, \mu)$ \cite{AStanislavsky17, AStanislavsky19}. The Laplace (L\'{e}vy) exponent  
\footnote{\label{12/02-f1} The Laplace (L\'{e}vy) exponent $\Psi(s)$ is related to the distribution of a nonnegative stochastic process $U(\xi)$ by $\langle \exp[-s U(\tau)] \rangle = \exp[-\tau \Psi(s)]$ where $\langle \cdot \rangle$ denotes the mean value.}
is denoted as $\Psi(\I\!\omega)$ and it is connected to the memory $\kappa(t)$ through the inverse Laplace transform, namely $\kappa(t) = \mathcal{L}^{-1}\{[\Psi(s)]^{-1}; t\}$ \cite[Eq. (23)]{AStanislavsky19}. Comparing Eq. \eqref{12/02-1} with Eq. \eqref{9/02-3} we obtain $\Psi(\alpha, \nu, \mu; a; s) = s \hat{k}(\alpha, \nu, \mu; a; s)$ as well as}
\begin{equation}\label{10/02-1}
{\kappa(\alpha, \nu, \mu; a; t) = \mathcal{L}^{-1}\{[s \hat{k}(\alpha, \nu, \mu; a; s)]^{-1}; t],} 
\end{equation}
{
From this relation \eqref{10/02-1} it turns out that the inhomogeneous integral equation \eqref{9/02-1} in the Laplace space reads $\hat{f}(s) = s^{-1} - B \hat{\kappa}(s) \hat{f}(s)$ with $\hat{f}(s) = \mathcal{L}[f(t); s]$.  Simple algebra gives
\begin{equation}\label{11/02-3}
\hat{f}(s) = \frac{[s \hat{\kappa}(s)]^{-1}}{[\hat{\kappa}(s)]^{-1} + B}.
\end{equation}
Eq. \eqref{10/02-1} enables one to set $[\hat{\kappa}(\alpha, \nu, \mu; a; s)]^{-1} = s \hat{k}(\alpha, \nu, \mu; B, a; s)$ and $[s \hat{\kappa}(\alpha, \nu, \mu; a; s)]^{-1} = \hat{k}(\alpha, \nu, \mu; B, a; s)$. When we insert these formulae into Eq. \eqref{11/02-3} then its inverse Laplace transform equals Eq. \eqref{20/03-1} and we can say that the inhomogeneous integral equation \eqref{9/02-1}, in which the quantity $r(t)$ in $\D f(t)/\D t = - r(t) f(t)$ is smeared and the homogeneous integro-differential equation \eqref{12/03-2}, where we smear the time derivative $\D f(t)/\D t$, leads to the same results. Thus, we can conclude that these two kinds of smearing give equivalent relations and it depends of our convenience which one to use. 
\begin{example}
For $a = 0$ we have $\hat{k}(\alpha, \nu, \mu; 0; s) = s^{\mu - 1}$  for which the relevant Laplace exponent $\Psi(s)$  equals to $s^{\mu}$. Hence, Eq. \eqref{10/02-1} yields to $\kappa(\alpha, \nu, \mu; 0; t) = t^{\mu - 1}/\Gamma(\mu)$ such that the inhomogeneous integral equation \eqref{9/02-1} can be written as
\begin{equation}\label{10/02-13}
f(\alpha, \nu, \mu; 0; t) = 1 - \tau^{\mu}\, [D^{- \mu}_{t} f(\alpha, \nu, \mu; 0; t)]\qquad \text{for} \qquad \mu\in(0, 1),
\end{equation}
where 
\begin{equation*}
(D^{- \eta}_{t} h)(t) = \frac{1}{\Gamma(\eta)} \int_{0}^{t} (t -\xi)^{\eta - 1} h(\xi)\D\xi, \qquad \eta\in(0, 1), 
\end{equation*}
is the Riemann-Liouville fractional integral. Eq. \eqref{10/02-13} can be transformed into the homogeneous integro-differential equation \eqref{12/03-2} by acting on it  the fractional derivative in the Riemann-Liouville sense $D^{\mu}_{t}$. Employing the semigroup property of the fractional derivative (integral) $D^{\eta}_{t}[D^{-\eta}_{t} h](t) = h(t)$ and the relation $D^{\eta}_{t} 1 = t^{-\eta}/\Gamma(1 - \eta)$ we arrive at the homogeneous integro-differential equation \eqref{12/03-2} 
\begin{equation*}%\label{10/02-14}
D_{t}^{\mu} f(\alpha, \nu, \mu; 0; t) - \frac{t^{-\mu}}{\Gamma(1-\mu)} = -\tau^{\mu} f(\alpha, \nu, \mu; 0; t), 
\end{equation*}
which LHS, using the relation between the fractional derivatives in the Caputo ${^{C\!}D_{t}}^{\eta}$ and the Riemann-Liouville $D_{t}^{\eta}$ senses \cite{IPodlubny99}
\begin{equation*}
(\,{^{C\!}D_{t}}^{\eta} h)(t) = (D_{t}^{\eta} h)(t) - \frac{t^{-\eta}}{\Gamma(1-\eta)}, \quad \eta\in(0, 1),
\end{equation*}
is rewritten as the commonly used equation for the Cole-Cole model involving the Caputo fractional derivative.
\end{example}}

\noindent
{\textbf{Remark 3.}
Due to Eq. \eqref{10/02-1} 
\begin{equation*}%\label{11/02-1}
s \hat{k}(\alpha, \nu, \mu; a; s) \, \hat{\kappa}(\alpha, \nu, \mu; a; s) = 1
\end{equation*}
the memory kernels $s \hat{k}(\alpha, \nu, \mu; a; s)$ and $\hat{\kappa}(\alpha, \nu, \mu; a; s)$ are the pair of the coupled quantities,   like, \textit{e.g.}, the Caputo fractional derivative and fractional integral in the case of the Cole-Cole model for which $s \hat{k}(\alpha, \nu, \mu; 0; s) = s^{\mu}$ and $\hat{\kappa}(\alpha, \nu, \mu; 0; s) = s^{-\mu}$. Thus, we can suspect that for the generalized fractional derivative, i.e. the integro-differential operator with the kernel $k(t)$, exists the associated with them the generalized fractional integral, i.e. the integral operator with the kernel $\kappa(t)$.}

\section*{Acknowledgments}

The authors express their gratitude to the anonymous referee whose suggestions and critical remarks caught their attention to the integral equations describing the relaxation processes and important references treating mutual relations between the evolution equations and stochastic approach to the relaxation processes. This enabled us not only to amend the paper but also to push forward our understanding of the relaxation phenomena and their mathematical description.       

K.G. and A.H. thank for the support provided to them by the Polish National Center for Science (NCN) under the research grant OPUS12 no.~UMO-2016/23/B/ST3/01714.~K. G. acknowledges support of the Polish National Agency for Academic Exchange (NAWA) provided to her under the Bekker Program, project no. PPN/BEK/2018/1/00184, as well as the warm hospitality of the ENEA Research Center, Frascati, Italy where the significant part of this research was done.

\appendix
\section{${h_{l/k, \lambda}(u, t) = \mathcal{L}^{-1}[s^{-\lambda} \E^{- u s^{l/k}}; t]}$}

The derivation of the  Meijer $G$ function representation of $h_{l/k, \lambda}(u, t)$ can be divided into three steps. {We start with}
\begin{equation*}
s^{-\lambda} \E^{-\xi s^{\alpha}} = \mathcal{L}[h_{\alpha, \lambda}(u, t); s].
\end{equation*}
In the next step, due to \cite[Theorem 1]{JSLew75} we calculate the Mellin transform of $s^{-\lambda} \exp(-\xi s^{\alpha})$ \footnote{\label{f2} The direct Mellin transform of $g(x)$ is equal to
\begin{equation*}%\label{17/05-2}
h^{\rm Mellin}(p) =\mathcal{M}[h(x); p] = \int_{0}^{\infty} x^{p-1} h(x) \D x, \qquad \text{where} \qquad h(x) = \mathcal{M}^{-1}[h^{\rm Mellin}(p); x] = \int_{L_{M}} x^{-p} h^{\rm Mellin}(p) \D p
\end{equation*}
is the inverse Mellin transform and $L_{M}$ denotes the Bromwich contour $\RE(p) > 0$.} 
{for} which {we use} the integral definition of gamma function. {That} yields to
\begin{equation*}
h^{\rm Mellin}_{\alpha, \lambda}(u, p) = [\Gamma(1-p)]^{-1} \int_{0}^{\infty} s^{- (p + \lambda)} \E^{- u s^{\alpha}} \D s = u^{\frac{\lambda}{\alpha} - \frac{1-p}{\alpha}}\, \frac{\Gamma(-\frac{\lambda}{\alpha} + \frac{1-p}{\alpha})}{\alpha \Gamma(1-p)}.
\end{equation*}
In the last step, we calculate the inverse Mellin transform of $h^{\rm Mellin}_{\alpha, \lambda}(u, p)$ for rational $\alpha = l/k$:
\begin{align*}
h_{\alpha, \lambda}(u, t) &= \frac{u^{\frac{k}{l}\lambda}}{t} \frac{k}{2\pi\!\I l}\int_{L_{M}} \Big(\frac{u^{k}}{t^{l}}\Big)^{\!-\frac{1-p}{l}}\; \frac{\Gamma(-\frac{k}{l} + k \frac{1-p}{l})}{\Gamma(1-p)} \D p \\ & = \frac{u^{\frac{k}{l}\lambda}}{t}  \frac{k}{2\pi\I} \int_{\tilde{L}_{M}} \Big(\frac{u^{k}}{t^{l}}\Big)^{\!-z}\; \frac{\Gamma(-\frac{k}{l}\lambda + k z)}{\Gamma(l z)} \D z,
\end{align*}
in which we set $p = 1 - l z$ and according to this changes we also changed in an appropriate way the Bromwich contour $\tilde{L}_{M}$. Applying the Gauss multiplication formula for gamma functions and the definition of the Meijer $G$ function \cite[Eq. (8.2.1.1)]{APPrudnikov-v3} we have 
\begin{align}\label{13/02-1}
h_{l/k, \lambda}(u, t) & = (2\pi)^{\frac{l-k}{2}} \Big(\frac{u}{k}\Big)^{\frac{k}{l}\lambda} \frac{\sqrt{l k}}{t}\, G^{\,k, 0}_{l, k}\Big(\frac{u^{k} l^{l}}{k^{k} t^{l}}\Big\vert {\Delta(l, 0) \atop \Delta(k, -k\lambda/l)}\Big) \\ \label{14/02-1}
& = \frac{\sqrt{k}\, l^{1/2-\lambda}}{(2\pi)^{(k-l)/2}}\, G^{\,k, 0}_{l, k}\Big(\frac{u^{k} l^{l}}{k^{k} t^{l}}\Big\vert {\Delta(l, \lambda) \atop \Delta(k, 0)}\Big)
\end{align}
with $\Delta(n, a)$ defined below Eq. \eqref{7/04-1}. Observing that $\Delta(l, 0)$ and $\Delta(k, -k\lambda/l)$ can be written as  $\Delta(l, \lambda) - \lambda/l$ and $\Delta(k, 0) - \lambda/l$, respectively, as well as applying \cite[Eq. (8.2.2.15)]{APPrudnikov-v3} we transform Eq. \eqref{13/02-1} into Eq. \eqref{14/02-1}. Comparing Eq. \eqref{14/02-1} for $\lambda = 1$ with Eq. \eqref{6/05-5} it can be seen  
\begin{equation*}
\varPhi_{l/k}(u, t) = h_{l/k, 1}(u, t).
\end{equation*}

We also find the series form of $h_{l/k, \lambda}(u, t)$. This purpose is realized by transforming the Meijer $G$ function in Eq. \eqref{13/02-1} into the finite sum of generalized hypergeometric function ${_{1+l}F_{k}}$ with the help of \cite[Eq. (8.2.2.3)]{APPrudnikov-v3}. Using to the obtained formula the series form of  ${_{1+l}F_{k}}$ we obtain 
\begin{equation}\label{13/02-2}
h_{\alpha, \lambda}(u, t) = u^{\frac{\lambda - 1}{\alpha}} \sum_{r=0}^{\infty} \frac{(-1)^{r}}{r! \Gamma(\lambda - \alpha r)} (t u^{-1/\alpha})^{\lambda - 1 -\alpha r}, \quad \alpha = l/k \in(0, 1),
\end{equation}
which for $\lambda = \beta - \alpha\gamma$ gives
\begin{equation*}
g_{\alpha, \beta}^{\gamma}(u, t) = u^{\gamma - \beta/\alpha} h_{\alpha, \beta - \alpha\gamma}(u, t).
\end{equation*}


\begin{thebibliography}{99}
\bibitem{10/02-f1}
The direct and inverse Laplace transforms are given as 
\begin{equation} 
\hat{h}(s) = \mathcal{L}[{h}(t); s] = \int_{0}^{\infty} \E^{-s t} {h}(t) \D t, \quad \text{and} \quad {h}(t) = \mathcal{L}^{-1}[\hat{h}(s); t] = \frac{1}{2\pi\I} \int_{L} \E^{s t} \hat{h}(s) \D s\nonumber
\end{equation}
where $t\in{{\mathbb R}}_{+}$, $s\in{\mathbb C}$ and $L$ is the Bromwich contour.
\end{thebibliography}
\end{document}